\documentclass[11pt,reqno,a4,onecolumn,twoside,final,spanish]{article}

\usepackage{amsthm,amsmath,amssymb,latexsym,amsfonts,amscd} 
\usepackage{latexsym}
\usepackage{url}
\usepackage[T1]{fontenc}
\usepackage{ae}
\usepackage{aecompl}
\usepackage[small,sc,hang,bf]{caption} 
\usepackage{graphicx,color,ae,fancyvrb}
\usepackage{pst-all}
\usepackage{subfig}
\usepackage{float} 
\usepackage{verbatim}
\usepackage{enumerate}
\usepackage{array}
\usepackage{longtable}
\usepackage{multicol}
\usepackage{cancel}
\newcommand*\chancery{\fontfamily{pzc}\selectfont}

\newlength\mytemplen
\newsavebox\mytempbox

\makeatletter
\newcommand\mybluebox{%
    \@ifnextchar[
       {\@mybluebox}%
       {\@mybluebox[0pt]}}

\def\@mybluebox[#1]{%
    \@ifnextchar[
       {\@@mybluebox[#1]}%
       {\@@mybluebox[#1][0pt]}}

\def\@@mybluebox[#1][#2]#3{
    \sbox\mytempbox{#3}%
    \mytemplen\ht\mytempbox
    \advance\mytemplen #1\relax
    \ht\mytempbox\mytemplen
    \mytemplen\dp\mytempbox
    \advance\mytemplen #2\relax
    \dp\mytempbox\mytemplen
    \colorbox{myblue}{\hspace{1em}\usebox{\mytempbox}\hspace{1em}}}

\makeatother

\usepackage{natbib}

\usepackage{hyperref}

\usepackage{geometry}
\usepackage{calc}
	
\makeatletter
\renewcommand{\section}{\@startsection{section}{1}{0pt}{-3ex plus -1ex minus 0ex}{2ex plus 0ex}{\bf}}
\makeatother

\makeatletter
\renewcommand{\subsection}{\@startsection{subsection}{1}{0pt}{-2ex plus -1ex minus 0ex}{2ex plus 0ex}{\bf}}
\makeatother

\theoremstyle{definition}

\theoremstyle{remark}

\begin{document}

\renewcommand{\tablename}{Tabla}
\renewcommand{\figurename}{Figura}
\noindent

\begin{flushleft}
\textsl {\chancery  Memorias de la Primera Escuela de Astroestad\'istica: M\'etodos Bayesianos en Cosmolog\'ia}\\
\vspace{-0.1cm}{\chancery  9 al 13 Junio de 2014.  Bogot\'a D.C., Colombia }\\
\textsl {\scriptsize Editor: H\'ector J. Hort\'ua}\\
\href{https://www.dropbox.com/sh/nh0nbydi0lp81ha/AACJNr09cXSEFGPeFK4M3v9Pa}{\tiny {\blue Material suplementario}}
\end{flushleft}



\thispagestyle{plain}\def\@roman#1{\romannumeral #1}



\begin{center}\Large\bfseries Una revisión a la teoría básica del CMB \end{center}
\begin{center}\normalsize\bfseries A review of the basic theory of CMB \end{center}

\begin{center}
\small
\textsc{Jorge H. Mastache de los Santos \footnotemark[1]}
\footnotetext[1]{Instituto de Física. Universidad Nacional Autónoma de México. E-mail: \url{mastache@fisica.unam.mx}}

\end{center}

\noindent\\[1mm]
{\small
\centerline{\bfseries Resumen}\\

La Cosmología está progresando a pasos agigantados gracias a la cantidad espectacular de datos observacionales que se obtienen tanto de los experimentos en tierra como satélites. Un papel fundamental es desempeñado por las observaciones del Fondo Cósmico de Microondas (CMB por sus siglas en inglés, \textit{Cosmic Microwave Background}), la cual nos proporciona la prueba observacional más directa de los inicios del Universo. Las observaciones de la temperatura y las anisotropías en el CMB han jugado un papel fundamental en la definición del modelo cosmológico. Esta contribución tiene como objetivo resumir algunos de los conceptos básicos que hay detrás de la física del CMB. La mayor parte de los ingredientes del modelo cosmológico estándar son poco conocidos en términos de la física fundamental, por efemplo, la materia oscura y la energía oscura. Se discute cómo las observaciones actuales abordan algunas de estas cuestiones.\\

{\footnotesize
\textbf{Palabras clave:}
Física, Astrofísica, Cosmología, Radiación de Fondo Cósmico.
\\
\noindent\\[1mm]
{\small
\centerline{\bfseries Abstract}\\

The cosmic microwave background (CMB) provides the most direct observational test of the early universe. The observations of the temperature anisotropies in the CMB have played a key role in defining the cosmological model. This contribution aims to summarize some of the basic concepts behind the physics of the CMB. Most of the ingredients of the standard cosmological model are poorly understood in terms of fundamental physics, for instance, dark matter and dark energy. We discuss how current observations addressed some of these issues.\\

{\footnotesize
}

\newpage
\section{Introducción}

Hace ya veinte años desde el descubrimiento histórico de las fluctuaciones en la temperatura de la radiación de fondo de microondas cósmico por el satélite COBE, \cite{Smoot:1992td}. De esa fecha hasta el día de hoy se ha posicionado dentro de la comunidad científica el llamado modelo cosmológico estándar. La medición de las fluctuaciones del CMB ha sido fundamental para establecer las bases de este modelo, han demostrado que el Universo está constituido por dos componentes no contenidas en el modelo estándar: la materia obscura (\emph{DM}, por sus siglas en inglés) y la energía oscura.

La teoría del CMB cubre una amplia gama de la astrofísica por lo que el propósito de esta revisión es establecer la física que hay detrás del CMB. Comenzaremos por revisar algunos de los conceptos visto en las presentaciones sin entrar en el detalle riguroso de las matemáticas, con mucho más enfasis en las ideas que en el álgebra. Para aquellos quienes les interese abordar con detalle los temas que aqui se han de tratar he de sugerirles algunos resumenes y trabajos claves en el desarrollo de las teoría de perturbaciones lineales del CMB, \cite{Samtleben:2007zz,Kamionkowski:1999qc,Seljak:1996is,Ma:1995ey}. De la misma manera suguiero leer los primeros capítulos del libro \cite{Lesgourgues:2003fe},  el cual contienen una presentación similar a la que abordé pero más detallada. Opcionalmente, el lector también puede profundizar en temas como los efectos de los neutrinos, polarización y lentes gravitacionales que no se abordan aquí.

\section{El CMB y el Modelo Cosmológico Estándard}
En el modelo cosmológico estándar, el modelo $\Lambda$CDM, que describe nuestro Universo a través de una métrica de Friedmann-Robertson-Walker en el contexto de relatividad general através de un espacio plano, homogéneo e isótropico con pequeñas fluctuaciones, del orden de $10{-5}$, en la densidad de energía. El Universo ha evolucionado a partir de una fase densa y caliente durante el cual la materia y la radiación estaban en equilibrio térmico. El CMB es la radiación reliquia debida a los fotones primigenios de esta primera fase y su existencia es una piedra angular del modelo del Big Bang. La radiación CMB ahora se ha \textit{enfriado} hasta una temperatura de 2.725 K, pero conserva un espectro de cuerpo negro casi perfecto. Los bariones y los leptones constituyen el 4.5\% de la densidad de energía actual y la materia oscura fría (CDM; como hipótesis pasa esencialmente sólo las interacciones gravitatorias y velocidades térmicas insignificantes) el 22\%. El restante 73\% se presenta en forma de energía oscura y es la causante actual de la expansión acelerada del Universo. La energía oscura no se entiende en absoluto a nivel físico pero fenomenológicamente se comporta como un fluido con una ecuación de estado $p = -\rho$, para una constante cosmológica $\Lambda$.

La planitud y la isotropía del Universo a gran escala se explican perfectamente por un período hipotético de expansión casi exponencial -inflación cósmica- en el Universo primitivo. Durante un período de tan sólo $10^{-32}$ segundos, el Universo se expandió en tamaño en una magnitud de $e^{60}$ veces. La inflación aún no se entiende en un nivel fundamental, pero puede ser caracterizado por modelos simples con la presencia de un campo escalar $\phi$ que evolución lentamente con un potencial de autointeracción $V(\phi)$. Una característica atractiva de inflación es que proporciona de forma natural un mecanismo causal para generar perturbaciones primordiales en la curvatura y ondas gravitacionales. Fluctuaciones cuánticas de pequeña escala de la métrica del espacio-tiempo, se expanden más allá del radio de Hubble durante la inflación para después ser las semillas de las perturbaciones cosmológicas clásicas que eventualmente evolucionan en el crecimiento de la estructura a gran escala.

\subsection{Anisotropías en la temperatura}
El CMB carga con las huellas de las perturbaciones primordiales a través de la anisotropías en su temperatura ($10^{-5}$). Cuando los fotones se desacoplan de los electrones el Universo se hizo transparente, la época de la recombinación, esto define la superficie de la última capa de dispersión, y las fluctuaciones espaciales en la densidad de energía del CMB, el potencial gravitatorio sobre esta superficil son las fuentes de las anisotropías de la temperatura en el CMB. Con un poco más detalle, las anisotropías a lo largo de una dirección en el tiempo esta dada por
\begin{equation}
   \Theta|_{obs} = (\Theta + \psi)|_{dec} + ({\hat n} \cdot {\overrightarrow v}_b)|_{dec} + \int_{\eta_{dec}}^{\eta_0}d \eta(\Theta^\prime + \psi^\prime).
\end{equation}

Cada término tiene una interpretación física simple: donde el primer entre paréntesis es el término referente al efecto Sachs-Wolfe, el cuál incluye el término de la temperatura $\Theta_0$ y el término del potencial gravitacional $\psi$ evaluados en la última capa de dispersión. El segundo término de la ecuación es el denominado efecto Doppler. El último término en la ecuación es el efecto Sachs-Wolfe integrado (ISW) que involucra la integral de las derivadas temporales de los potenciales $\theta$ y $\psi$; si un pozo de potencial graviatacional es cada vez menos profundo en el tiempo (como ocurre durante la dominación de energía oscura), la energía de los fotones tienen un corrimiento neto hacia el azul al cruzar el pozo de potencial. En la práctica, alrededor de 10\% de los fotones se dispersó después de la última capa de dispersión por efectos como el ISW.

La peque\~na amplitud de las anisotropías de la temperatura implica que se pueden calcular con la teoría de perturbación lineal con mucha precisión. Las fluctuaciones de la última capa de dispersión son, por tanto, la versión lineal de las perturbaciones primordiales. A escalas grandes (en comparación con el radio de Hubble) al fin de la recombinación, sólo la gravedad es importante, pero para escalas menores la física de las oscilaciones acústicas del plasma primordial y difusión de fotones empiezan a dominar. La atración gravitacional tenderá a incrementar las perturbación con densidad positiva, pero la creación de oscilaciones acústicas en el plasma es constrarestado por la presión de los fotones. 

\begin{figure}[h!]
  \centering
  \includegraphics[width=0.8\textwidth]{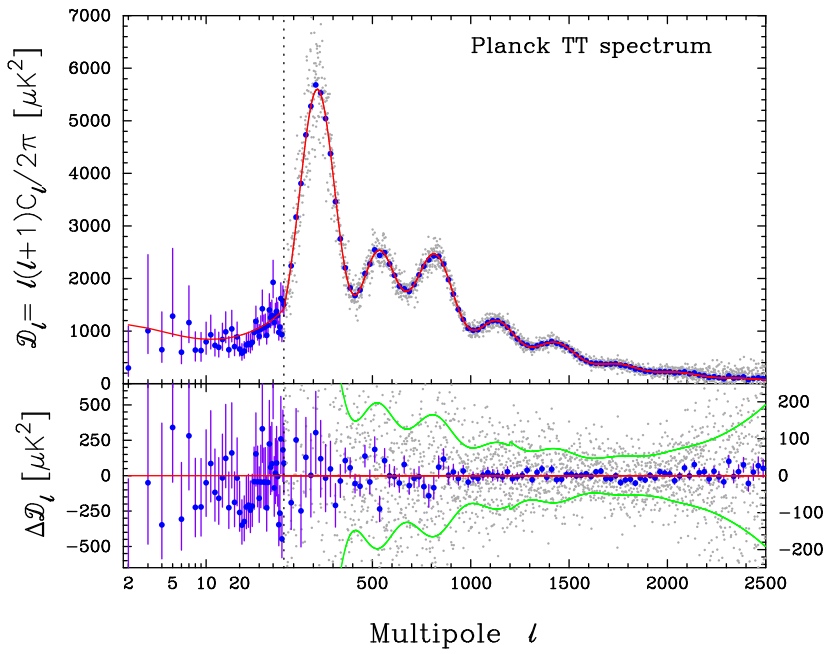} 
  \caption {Espectro de potencias pobservado por el satélite Planck, \cite{Ade:2013zuv}.}
  \label{fig1}
\end{figure}

La figura \ref{fig1} muestra el espectro de potencia predicho por la teoría, $D_l$. El espectro de potencia es la varianza de los múltiplos $\Theta_{lm}$ en una expansión  de armónicos esféricos $\Theta = \sum_{lm}\Theta_{lm}Y_{lm}$. El índice multipolar $l$ corresponde aproximadamente a las anisotropías a la escala de $180^{\circ}/l$. El comportamiento de la curva a grandes escalas se debe solamente a la combinación de las anisotropías primarias y el efecto Sachs-Wolfe integrado ya que ha estas escalas las oscilaciones acústicas no afectan el comportamiento del espectro de potencias. En las escalas intermedias, tenemos picos acústicos. Por último, en las escalas más pequeñas la espectro de potencias decae rápidamente. Este decaimiento se debe a la amortiguación de las perturbaciones ya que los fotones tuvieran tiempo de difundirse fuera de las sobredensidades al momento de la última capa de dispersión, provocando así un amortiguamiento las oscilaciones acústicas. Este proceso imprime otra escala, la escala de la difusión, en el CMB.

El panorama descrito anteriormente se confirma espectacularmente por las mediciones de las anisotropía de la temperatura por los satélites, desde el satélite COBE, WMAP y el más reciente Planck, \cite{Ade:2013zuv}. La teoría sólo nos permite predecir las propiedades estadísticas de las perturbaciones primordiales. En los modelos más simples de inflación, mantiene la suposición que las propiedades estadísticas de espectro de postencia son gaussiana, éstas caracterizan el espectro de potencia de maneras particulares. Verificar que las perturbaciones provengan de gaussianidades primordial depende de las mediciones cuidadosas de la estadística de las anisotropías del CMB. Por esta razón, la investigación observacional del CMB durante los últimos 20 a\~nos ha sido la de obtener estimaciones precisas del espectro de potencia y comparar estos con los modelos teóricos.

Teniendo en cuenta que la física del CMB está tan bien entendidas, el modelo cosmológico estándar puede ser probada de forma muy precisa y sus parámetros se determinan con gran precisión (ver \cite{Ade:2013zuv}). 

En este resumen y por razones de brevedad, destacaremos varios ejemplos:

\textit{Espectro de potencia primordial}: La morfología general del espectro de potencia del CMB depende de la parametrización del espectro primordiales, la cuál suele utilizarse una ley de potencias, $P_R(k) \propto k^{n_s-1}$, donde se ha medido que $n_s = 0.9616 \pm 0.0094$. Esto es muy consistente con la predicción de inflación, el cual predice que $n_s = 1$. Una desviación de la invarianza de escala del espectro de potencia ($n_s \neq 1$) se detecta a ordenes de $3\sigma$ y ofrece limitaciones importantes en la dinámica del potencial que regula inflación, $V(\phi)$.

\textit{Densidades de Materia}: Las alturas relativas de los picos acústicos están influenciados por las densidades de bariones y la CDM. Por ejemplo, el aumento de la fracción de bariones añade la inercia al fluido primigenio pero sin añadir presión a el plasma, lo que reduce el módulo de compresibilidad del fluido. Esto aumenta la sobredensidad en el punto medio de las oscilaciones acústicas e impulsar los picos de compresión (los picos impares). Las constriciones de la densidad de los bariones $\Omega_b h^2 = 0.02207 ± 0.00033$, se han realizado gracias a la medición de las alturas relativas de los picos acústicos, y que son consistentes con las constricciones impuestas del Big Bang nucleosíntesis. Del mismo modo, la densidad de la CDM se ha restringido en $\Omega_c h^2 = 0.1196 ± 0.0031$. Esto proporciona evidencia de la necesidad de oscura no-bariónica, cuyo requerimento se complementa con las mediciones independientes tales como las curvas de rotación de galaxias y cinemática de c\'umulo de galaxias.

\textit{Curvatura}: La escala angular de los  picos acústicos, $r_s/d_A$, se mide con mucha precisión. En los modelos estándar, las densidades de materia determinadas a partir de las alturas relativas de los picos ayuda a determinar el valor de $r_s$, permitiendo así una medición precisa de la distancia angular en el momento de la última dispersión. Esta distancia es muy sensible a la curvatura espacial, pero que a su vez depende del valor del parámetro de Hubble, $H_0$ ó equivalente, de la densidad de energía oscura. Esto conduce a una degeneración en el cual los modelos con las mismas densidades de energía (para corrimientos al rojo grandes), el mismo espectro de potencia primordial, y la misma distancia angular en el momento de la última capa de dispersión dan espectros de potencia casi idénticos. La degeneración se puede romper mediante la adición de otras mediciones astrofísicas como la distancia a las supernovas tipo Ia, la constante de Hubble, la distancia angular a corrimientos al rojo bajos. Un temprano ejemplo importante de la determinación de la curvatura con este conocimiento fue gracias a BOOMERanG \cite{deBernardis:2000gy}, quienes midieron  con precisión el primer pico acústico, el equipo fue capaz de establecer que el espacio era plano con una exactitud del 10\%. Mediciones más recientes son consistentes que el Universo es plano con una exactitud del 0.2\% \cite{Ade:2013zuv}, apoyando fuertemente una de las principal es predicciones de inflación.

\section{Conclusiones}
Espero que esta breve introducción a las perturbaciones cosmológicas sea lo suficientemente clara, y que el lector se sienta entusiasmado a profundizar más en el tema. Algunos lectores podrían lamentar la falta del detalle matemático, que quedan fuera de los alcances de este resumen, para ello he de sugerir el trabajo de  \cite{Ma:1995ey},  el cual contiene una presentación clara de las ecuaciones de evolución de la métrica y las ecuaciones de movimiento de los fluidos en las normas más populares en la literatura (Newtoniana y Síncrona), que son las ecuaciones fundamentales para el análisis de las perturbaciones cosmológicas.

\section{Agradecimientos}
Agradezco enormemente el apoyo brindado por la Fundaci\'on Universitaria Los Libertadores, en particular a los organizadores Javier Hort\'ua Orjuela y Carolina Cabrera por darme la oportunidad de dar el taller y brindar un ambiente de amabilidad. Estoy también agradecido a los participantes al taller quienes mostraron gran entusiasmo durante la charla.

\renewcommand{\refname}{Bibliografía}
\bibliographystyle{harvard}
\bibliography{Mastache}

\end{document}